\documentstyle[epsf]{cjp3}

\begin{document}

\title{Thermal hard-photons probing multifragmentation\\
in nuclear collisions around the Fermi energy}

\authori{{D.G. d'Enterria}$^{a,b,*}$) and G. Mart\'{\i}nez $^{a,}$\footnote
{Present address: SUBATECH, BP 20722, 44307 Nantes Cedex 3, France}
\\for the TAPS collaboration}

\addressi{$^a$GANIL, BP~5027, 14076 Caen Cedex 5, France\\
$^b$Grup de F\'{\i}sica de les Radiacions, Universitat Aut\`onoma de Barcelona, \\08193 Cerdanyola, Catalonia
}

\authorii{}       
\addressii{}
\authoriii{}      
\addressiii{}     

\headtitle{Thermal hard-photons probing multifragmentation$\ldots$}
\headauthor{David d'Enterria}

\evidence{A}
\daterec{XXX} 
\cislo{0}  \year{2000}
\setcounter{page}{1}

\maketitle

\begin{abstract}
Hard-photon (E$_{\gamma}\,>$ 30 MeV) emission issuing from
proton-neutron bremsstrahlung collisions is investigated in four
different heavy-ion reactions at intermediate bombarding energies
($^{36}$Ar+$^{197}$Au, $^{107}$Ag, $^{58}$Ni, $^{12}$C at 60{\it A}
MeV) coupling the TAPS photon spectrometer with two
charged-particle multidetectors covering more than 80\% of the
solid angle. The hard-photon spectra of the three heavier targets
result from the combination of two distinct exponential
distributions with different slope parameters, a result which
deviates from the behaviour expected for hard-photon production
just in first-chance proton-neutron collisions. The thermal origin
of the steeper bremsstrahlung component is confirmed by the
characteristics of its slope and angular distribution. Such thermal
hard-photons convey undisturbed information of the thermodynamical
state of hot and excited nuclear systems undergoing
multifragmentation.
\end{abstract}

\section{Introduction}

The production of elementary particles not present in the nucleus
as real particles (such as photons, mesons or dileptons) has
attracted much interest in heavy-ion physics at intermediate
bombarding energies (20{\it A} MeV $\leq$ $E_{lab}$ $\leq$ 100{\it
A} MeV) since they are precise probes of the phase-space evolution
of nucleus-nucleus collisions \cite{enterria:Cassing90}. Among the
experimental probes, photons are leading observables because, at
variance with particles and nuclear fragments, after their
production they do not suffer any final-state (either Coulomb or
strong) interaction with the surrounding medium, giving a faithful
image of the emission source. Above E$_{\gamma}$ = 30 MeV, a bulk
of experimental and theoretical evidences indicate that photons are
mainly produced during the first instants of the reaction in
individual proton-neutron bremsstrahlung ($pn\gamma$) collisions
within the participant zone
~\cite{enterria:Cassing90,enterria:Nifenecker90}. Hard-photons
reveal, hence, the preequilibrium conditions existing in the
initial high-compression phase of the reaction. Recent experimental
results collected with a second-generation of photon spectrometers
such as the ``Two Arm Photon Spectrometer"
TAPS~\cite{enterria:Novotny91} coupled to charged-particle
multidetectors covering a large fraction of the available
phase-space, have allowed a more precise and exclusive analysis of
hard-photon energy spectra, angular distributions and two-photon
correlation functions in a wide photon energy range, with high
statistics data sets and for different projectile-target
combinations ~\cite{enterria:Martinez95,enterria:Marques95}. These
new measurements have pointed out that, aside from the pure
first-chance (or ``direct") $pn\gamma$ emission scenario, an
additional hard-photon softer component (referred to as ``thermal"
hard-photons) shows up accounting for up to a third of the total
hard-photon yield. Those thermal hard-photons are emitted in
secondary $pn\gamma$ collisions in a later stage of the reaction
when part of the available energy has been thermalized among
intrinsic degrees of freedom
\cite{enterria:Martinez95,enterria:Schutz97}.
\\\\
The interest of such an observation relies on the fact that thermal
hard-photons would constitute a singular probe of the intermediate
dissipative stages of the reaction where the process of nuclear
fragmentation is supposed to take place, giving hints on both the
time-scale of this process and, eventually, on the thermodynamical
state (temperature, density, degree of thermalization ...) of the
fragmenting source(s). This information is fundamental to
understand the mechanism which drives the copious production of
intermediate-mass-fragments (IMF, $3\,\leq\,Z\leq\,20$)
experimentally observed in intermediate-energy heavy-ion
collisions. Indeed, in the present-day there is no general
agreement among the different experimental results and the
different statistical ~\cite{enterria:Bondorf95,enterria:Gross97}
or dynamical ~\cite{enterria:Aichelin91} theoretical models
regarding the underlying physical mechanism behind such
multifragmentation process (see e.g. ~\cite{enterria:Hirschegg99}).
The key issue in these studies is whether or not an equilibrated
system is created in the course of the collision. This is essential
to extract valid information on the thermodynamical properties of
the highly excited nuclei produced in such reactions, and
ultimately to signal the existence of a liquid-gas phase transition
in finite nuclear systems.
\\\\
To confirm the existence of thermal hard-photon emission and to
study thoroughly this process, two experimental campaigns of the
TAPS collaboration have been carried out in 1997 and 1998 at the
KVI and GANIL \cite{enterria:Orte00} facilities coupling, for the
first time, a photon spectrometer with two different
charged-particle multidetectors covering more than 80\% of 4$\pi$.
Here we report some first results of hard-photon and fragment
production in the $^{36}$Ar+$^{197}$Au, $^{107}$Ag, $^{58}$Ni,
$^{12}$C reactions at 60{\it A} MeV studied at KVI.
\\\\
The paper is organized as follows. Section 2 is devoted to the
description of the experimental setup. The first inclusive and
exclusive experimental results obtained and their interpretation
are shown in Section 3. A summary is finally given in Section 4.

\section{Experimental setup}

\begin{figure}
 \begin{center}
 \mbox{\epsfxsize=9.0cm \epsfbox{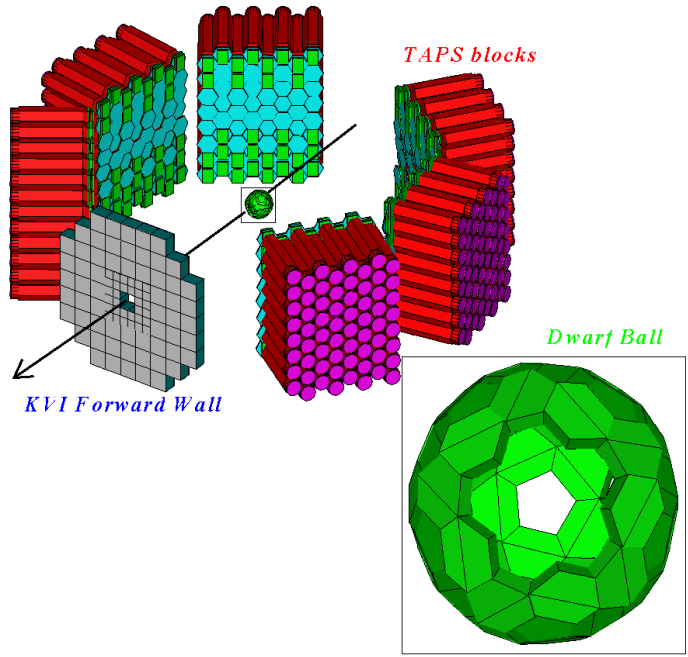}}
 \end{center}
 \caption{\it GEANT layout of the detection system composed of TAPS electromagnetic
 calorimeter, and Dwarf-Ball and Forward Wall charged particle multidetectors.}
 \label{fig:enterria:1}
\end{figure}

The experiment was performed end of 1997 at the KVI (Groningen)
facility using the $^{36}$Ar beams delivered by the K = 600
superconducting AGOR cyclotron at 37.1 MHz. Four systems were
studied, $^{36}$Ar+$^{197}$Au, $^{107}$Ag, $^{58}$Ni, $^{12}$C at
60{\it A} MeV, with different beam intensities (ranging from 3.0 nA
up to 12.5 nA) and target thicknesses (from 1 mg/cm$^2$ to 18
mg/cm$^2$). The goals of the campaign were to study exclusively
photon production in a wide energetic domain, ranging from
statistical Giant-Dipole-Resonance gamma-rays (10 MeV $\leq
E_{\gamma} \leq $ 20 MeV) to very energetic hard-photons (up to
E$_{\gamma} \approx $ 200 MeV), in coincidence with multi-fragment
emission processes for different heavy-ion systems. For that
purpose, we used concurrently a wide dynamic range photon
spectrometer, together with two phoswich multidetectors for
light-charged-particles (LCP, Z $\leq$ 2) and
intermediate-mass-fragments (IMF, 3 $\leq$ Z $\leq$ 10) detection
(fig. \ref{fig:enterria:1}). TAPS electromagnetic calorimeter
comprising 384 BaF$_2$ scintillation modules in a six-block
configuration and covering a solid angle of about 15\% of $4\pi$
was used to measure the double differential cross section
$d\sigma/d\Omega_\gamma dE_\gamma$ for photons of 10 MeV
$\leq\,E_{\gamma}\,\leq$ 200 MeV. Photons were separated from
charged particles and neutrons on the basis of pulse-shape
analysis, time-of-flight and TAPS charged-particle veto information
\cite{enterria:Marques95b}.

The Washington University ``Dwarf-Ball" (DB) \cite{enterria:Stra90}
(composed of 64 BC400-CsI(Tl) phoswich telescopes in the angular
range $32^\circ < \theta < 168^\circ$ covering around 76\% of
4$\pi$) and the KVI's ``Forward Wall" (FW) hodoscope
\cite{enterria:Leeg92} (92 NE102A-NE115 $\Delta$E-E phoswich
detectors in the forward angular range $2.5^\circ < \theta <
25^\circ$ with a geometrical acceptance of 4\% of 4$\pi$) allowed
us to identify isotopically the different LCP produced ($p$, $d$, $t$,
$^3$He and $\alpha$) and the charge of the IMFs up to that of the
projectile by means of pulse-shape analysis techniques.
\\
The first reaction, $^{36}$Ar+$^{197}$Au, was thoroughly studied
with the whole experimental setup (TAPS, DB and FW, see fig.
\ref{fig:enterria:1}) at low beam intensities; whereas the rest of
the reactions were studied with a reduced setup (TAPS and FW)
during a second set of high counting-rate runs.

\section{Experimental results}

Using the setup shown in fig. \ref{fig:enterria:1}, more than
1.1$\cdot 10^6$ hard-photons for the four mentioned systems were
collected out of 180$\cdot 10^6$ recorded events taking into
account the different defined triggers. When dealing with
nucleus-nucleus reactions, a first analysis of the experimental
results usually considers inclusive (i.e. $A+B\rightarrow
\gamma+X$, regardless of the particular final exit-channel)
observables such as the shape of the photon spectra and the angular
distributions (sections 3.1 and 3.2). Results from exclusive
reactions, i.e. from selected $\gamma$-particles coincidence
exit-channels, will be presented in section 3.3.

\subsection{Hard-photon energy spectra}

\begin{figure}
 \begin{minipage}[t]{.495\linewidth}
   \center\epsfxsize= 6.9cm \epsfbox{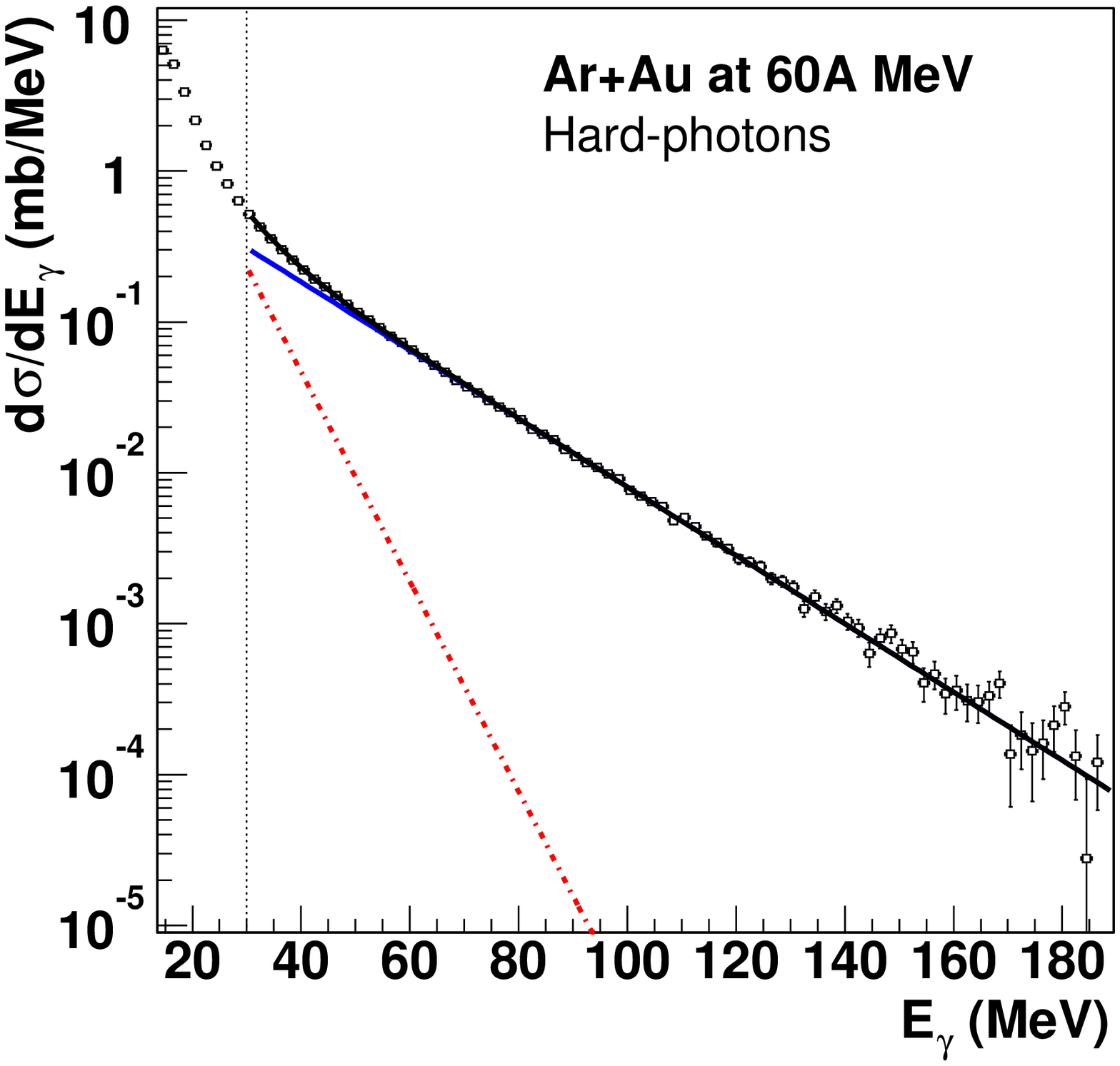}
 \end{minipage}
 \begin{minipage}[t]{.495\linewidth}
   \center\epsfxsize= 6.9cm \epsfbox{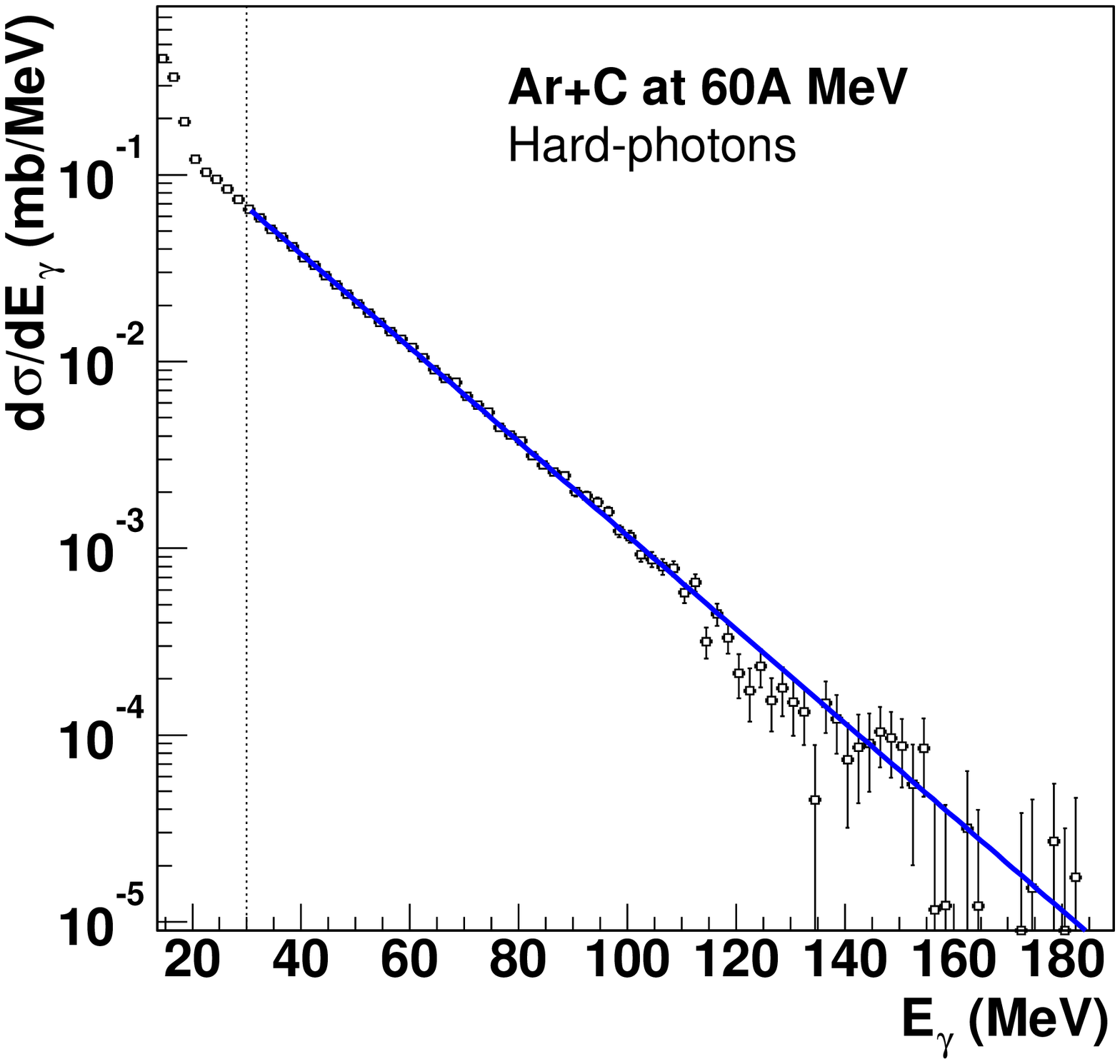}
 \end{minipage}
 \caption{\it Experimental hard-photon (E$_{\gamma}\,>$ 30 MeV) spectra
measured in the $NN$ center-of-mass frame for the heaviest
($^{36}$Ar+$^{197}$Au, left side) and the lightest
($^{36}$Ar+$^{12}$C, right side) systems and fitted, according to
equation (\ref{eq:enterria:1}), to the sum of two exponential
distributions: a direct (solid line) and a thermal one (dashed
line).}
 \label{fig:enterria:2}
\end{figure}

In fig. \ref{fig:enterria:2} we show the photon energy spectra in the
$NN$ CM frame for the heaviest ($^{197}$Au) and lightest ($^{12}$C)
targets after correcting for the TAPS response function and after
subtraction of the cosmic and 2-$\gamma$ $\pi^0$-decay backgrounds.
A steeper slope in the region $E_\gamma$ = 30 - 60 MeV with respect
to the flatter exponential fall-off describing the high-energy part
of the spectra above $E_\gamma$ = 60 MeV is clearly observed in the
heaviest $^{36}$Ar+$^{197}$Au system (as well as in the $^{107}$Ag
and $^{58}$Ni targets, not shown here) but it is not present in the
lightest $^{36}$Ar+$^{12}$C reaction. In the three heavier systems,
this excess of hard-photons clearly hinders the fit with a single
exponential of the spectra above 30 MeV as it has been commonly
done in hard-photon studies \cite{enterria:Nifenecker90}. To
properly describe the measured spectra for the heavier systems we
have, therefore, applied a sum of two exponential distributions
corresponding, as proposed by
\cite{enterria:Martinez95,enterria:Schutz97}, to a ``direct"
(coming from first-chance $pn\gamma$ collisions) and a ``thermal"
(secondary $pn\gamma$) hard-photon component respectively, with
their corresponding weights:
\begin{equation}
\frac{d\sigma}{d E_\gamma}\,=\,K_d\:e^{-E_\gamma/E_0^d}\,+\,K_t\:e^{-E_\gamma/E_0^t}
\label{eq:enterria:1}
\end{equation}

The slopes of the direct component ($E_0^d\,\approx$ 20 MeV) are
between two to three times larger than the thermal ones
($E_0^t\,\approx$ 6 - 9 MeV) and the contribution of thermal
hard-photons represents a 15\% - 20\% of the total hard-photon
yield for the three heavier targets (Table \ref{tab:enterria:1}).
In the small $^{36}$Ar+$^{12}$C projectile-target combination, the
resulting zone of participant nuclear matter does not have the
necessary volume to achieve sufficient stopping and consequent
thermalization and, hence, for that system pure direct
bremsstrahlung clearly dominates the whole photon emission above
$E_\gamma$ = 30 MeV.

\begin{table}
\begin{center}
\caption{\it  Ratios of thermal to total intensities, as well as direct
and thermal slopes for the hard photons measured in the
$^{36}$Ar+$^{197}$Au, $^{107}$Ag, $^{58}$Ni, $^{12}$C reactions at
60{\it A} MeV. The available energies in the lab and in the AA CM
frames for each reaction are also reported.}
\begin{tabular}{cccccc} \hline\hline
\small{System} & \small{$I_{t}/I_{tot}$} & \small{$E_0^{d}$  (MeV)} & \small{$E_0^{t}$ (MeV)} & \small{$E_{Cc}^{lab}$ ({\it A}MeV) } & \small{$E_{Cc}^{AA}$ ({\it A}MeV)}\\ \hline
\small{$^{36}$Ar+$^{197}$Au} & \small{0.19 $\pm$ 0.01} & \small{20.2 $\pm$ 1.2}  &  \small{6.2 $\pm$ 0.5}  & \small{55.5}  & \small{ 7.3} \\
\small{$^{36}$Ar+$^{107}$Ag} & \small{0.15 $\pm$ 0.01} & \small{20.1 $\pm$ 1.3}  &  \small{6.1 $\pm$ 0.6}  & \small{57.0}  & \small{10.7} \\
\small{$^{36}$Ar+$^{58}$Ni}  & \small{0.20 $\pm$ 0.01} & \small{20.9 $\pm$ 1.3}  &  \small{8.8 $\pm$ 0.8}  & \small{58.1}  & \small{13.7} \\
\small{$^{36}$Ar+$^{12}$C}   & \small{0.00 $\pm$ 0.05} & \small{18.1 $\pm$ 1.1}  &  \small{0.0 $\pm$ 0.5}  & \small{59.5}  & \small{11.1} \\ \hline\hline
\end{tabular}
\label{tab:enterria:1}
\end{center}
\end{table}

\begin{figure}
 \begin{center}
 \mbox{\epsfxsize=9.5cm \epsfbox{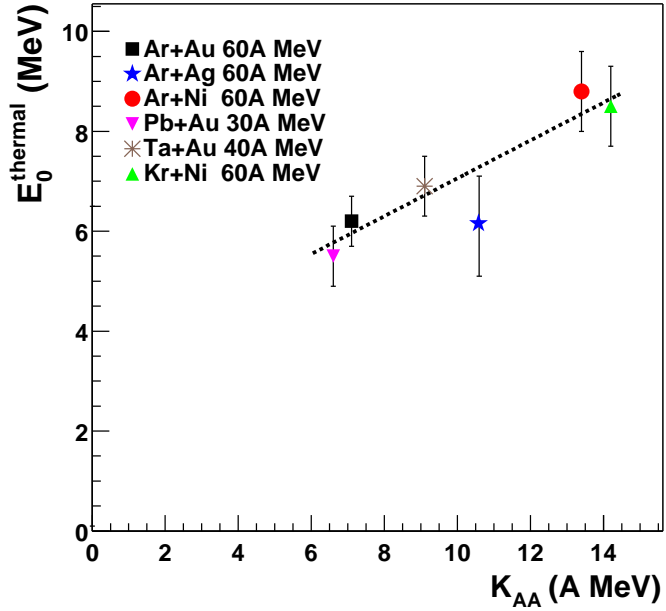}}
 \end{center}
 \caption{\it Thermal hard-photon slope parameters, $E_0^t$, as a function of the
 (Coulomb-corrected) nucleus-nucleus center-of-mass energy, $K_{AA}$, for the six
 systems studied in the TAPS campaign at KVI in 1997 and in the 1992 TAPS campaign
 at GANIL \cite{enterria:Schutz97}. The dashed line is a linear fit to the data.}
  \label{fig:enterria:3}
\end{figure}

The slope parameters of the direct component follow the known
linear dependence with the projectile energy per nucleon in the
laboratory frame as expected for pre-equilibrium emission in
first-chance $NN$ collisions~\cite{enterria:Nifenecker90}. The
thermal hard-photon slopes, at variance, scale with the available
energy in the nucleus-nucleus center-of-mass (fig.
\ref{fig:enterria:3}), suggesting that they are emitted at a later
stage of the reaction when the incident kinetic energy has been
dissipated into internal degrees of freedom in the $AA$ system.


\subsection{Hard-photon angular distributions}

The interpretation of the second component of the hard-photon
spectrum as being emitted during later stages of the reaction is
confirmed by the study of the Doppler-shifted laboratory angular
distributions. The angular distribution of bremsstrahlung photons
emitted in heavy-ion reactions can be parametrized, in the frame of
the emitting source, by an isotropic and an anisotropic dipolar
term~\cite{enterria:Nifenecker90}. The measured angular
distributions in the laboratory system contain an additional
Lorentz-boost term,
$Z\,=\,\gamma_{S}\,(1\,-\,\beta_{S}\,\cos\theta_{lab})$, depending
on the velocity $\beta_{S}$ of the moving photon source:

\begin{equation}
\left(\frac{d\sigma}{d\Omega}\right)_{lab}\,=\,
\frac{K}{Z(\beta_{S})^2}\left[1\,-\,\alpha\,+\,\alpha\,\frac{\sin^2\theta^{lab}_\gamma}{Z(\beta_{S})^2}\right]\,E_0^{d}\, e^{-30[\mbox{\scriptsize{MeV}}]\cdot Z(\beta_{S})/E_0^{d}}
\label{eq:enterria:2}
\end{equation}

\begin{figure}
 \begin{center}
 \mbox{\epsfxsize=9.5cm \epsfbox{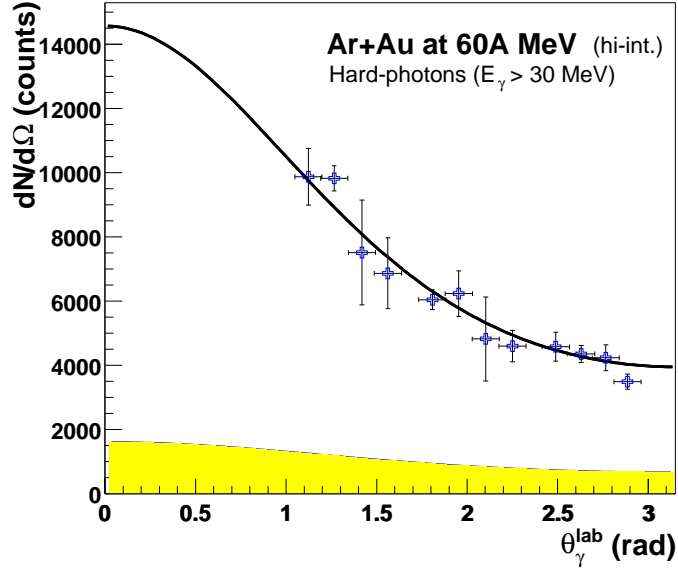}}
 \end{center}
 \caption{\it Experimental angular distribution in the lab for hard-photons
($E_{\gamma}>$ 30 MeV) measured in the system $^{36}$Ar+$^{197}$Au
during the high counting-rate runs and fitted according to equation
(\ref{eq:enterria:3}). The hatched region indicates the estimated
contribution of thermal hard-photons, emitted isotropically from a
source moving with $\beta_{AA}$.}
\label{fig:enterria:4}
\end{figure}

Above E$_{\gamma}$ = 60 MeV, the photon spectra are dominated by
``pure" direct hard-photons for all four systems and their angular
distributions are very well described with equation
(\ref{eq:enterria:2}) assuming an emission from a source moving
with the nucleon-nucleon center-of-mass velocity
($\beta_{S}\,\approx\,\beta_{NN}$ = 0.17). However, the average
source velocities obtained from the laboratory angular
distributions of the hard-photons with energies between 30 MeV and
45 MeV emitted in the three heavier reactions, are a factor 10\% to
25\% lower ($\beta_{S}\approx 0.14$). The presence in this photon
energy range of an important second bremsstrahlung component,
issuing from later nucleon-nucleon collisions taking place within a
slowly recoiling residue, induces this lowering of the average
velocity of the hard-photon moving source. As a matter of fact, the
hard-photon angular distributions above 30 MeV (fig.
\ref{fig:enterria:4}) can be well reproduced with the distribution
expected for the emission from a first-chance source with slope
parameter $E_0^d$ moving with $\beta_S^d\,\approx\,\beta_{NN}$ plus
a second-chance isotropic source with slope parameter $E_0^t$ and
$\beta_S^t\,\approx\,\beta_{AA}$ (the ratios of thermal to direct
intensities being fixed by those obtained from the energy spectra).
Such an expression reads:

\begin{equation}
\left(\frac{d\sigma}{d\Omega}\right)_{lab}\,=\,\frac{K}{Z^2}\left[1\,-\,\alpha\,+\,\alpha\,
\frac{\sin^2\theta_{\gamma}^{lab}}{Z^2}\right]\,E_0^{d}\,e^{-30[\mbox{\scriptsize{MeV}}]\cdot Z/E_0^{d}}
\,+\, \frac{K^{'}}{Z^{'2}}\,E_0^{t}\,e^{-30[\mbox{\scriptsize{MeV}}]\cdot Z^{'}/E_0^{t}}
\label{eq:enterria:3}
\end{equation}

where $E_0^{d,t}$ are the slope parameters of the direct and
thermal components in the source frame respectively, and
$Z\,=\,1/\sqrt{1-(\beta_S^{d})^2}\;\cdot(1-\beta_S^{d}\,\cos\theta_\gamma^{lab})$
and
$Z^{'}\,=\,1/\sqrt{1-(\beta_S^{t})^2}\;\cdot(1-\beta_S^{t}\,\cos\theta_\gamma^{lab})$
are the Lorentz-boost factors corresponding to the direct and
thermal source moving frames, $K,\,K^{'}$ are the normalization
factors associated to the direct and thermal intensities $I_{d}$
and $I_{t}$ respectively, and $\alpha$ is the weight of the dipole
component (only for the direct component). The values of the two
sources velocities $\beta_S^{d}$ and $\beta_S^{t}$ as well as the
intensity of the thermal component obtained with such a fit of the
laboratory angular distributions are summarized in table
\ref{tab:source velocities final}.

\begin{table}
\begin{center}
\caption{\it Direct and thermal hard-photon ($E_\gamma\,>$ 30 MeV) source
velocities obtained from a fit of hard-photon laboratory angular
distributions to equation (\ref{eq:enterria:3}).}
\label{tab:source velocities final}
\vspace{5mm}
\begin{tabular}{lccccc} \hline\hline
Reaction                  & $\beta_S^{d}$   & $\beta_S^{t}$   & $\alpha$        & $I_{t}/I_{tot}$ & $\chi^2/\nu$ \\ \hline
$^{36}$Ar+$^{197}$Au & 0.17 $\pm$ 0.02 & 0.06 $\pm$ 0.01 & 0.0 $\pm$ 0.0   & 19\% $\pm$ 2\%  &  1.7\\
$^{36}$Ar+$^{107}$Ag & 0.18 $\pm$ 0.01 & 0.09 $\pm$ 0.01 & 0.0 $\pm$ 0.0   & 19\% $\pm$ 2\%  &  0.3\\
$^{36}$Ar+$^{58}$Ni  & 0.18 $\pm$ 0.03 & 0.14 $\pm$ 0.02 & 0.0 $\pm$ 0.0   & 20\% $\pm$ 2\%  &  0.5\\
$^{36}$Ar+$^{12}$C   & 0.20 $\pm$ 0.01 & 0.00 $\pm$ 0.05 & 0.25 $\pm$ 0.05 &  0\% $\pm$ 2\%  &  1.1\\ \hline\hline
\end{tabular}
\end{center}
\end{table}

This result is consistent with direct hard-photons being emitted
from the nucleon-nucleon center-of-mass, and thermal hard-photons
being emitted isotropically from a source moving with the
nucleus-nucleus center-of-mass velocity.

\subsection{Exclusive results}

By selecting specific centrality channels by means of the measured
charged-particle multiplicity, we have been able to study as a
function of impact-parameter the yield of different types of
photons emitted in the $^{36}$Ar+$^{197}$Au reaction:

\begin{enumerate}
\item (Mainly) Giant-Dipole-Resonance (GDR) photons, defined as
photons with energies in the range 10 MeV $<$ E$_{\gamma}\,<$ 22 MeV,
\item ``Mixed"\footnote{According to the ratio of intensities of the
thermal component with respect to the total hard-photon yield above
30 MeV in the $^{36}$Ar+$^{197}$Au system (table
\ref{tab:enterria:1}), the relative proportion of thermal to direct
bremsstrahlung photons in the range 30 MeV $<$ E$_{\gamma}\,<$ 45
MeV is 40:60 (thus the term ``mixed" for this second selected type
of photons).} thermal and direct hard-photons, in the range 30 MeV
$<$ E$_{\gamma}\,<$ 45 MeV,
\item ``Pure" direct hard-photons (i.e. photons with E$_{\gamma}>$ 60 MeV).
\end{enumerate}

\begin{figure}
 \begin{center}
 \mbox{\epsfxsize=10cm \epsfbox{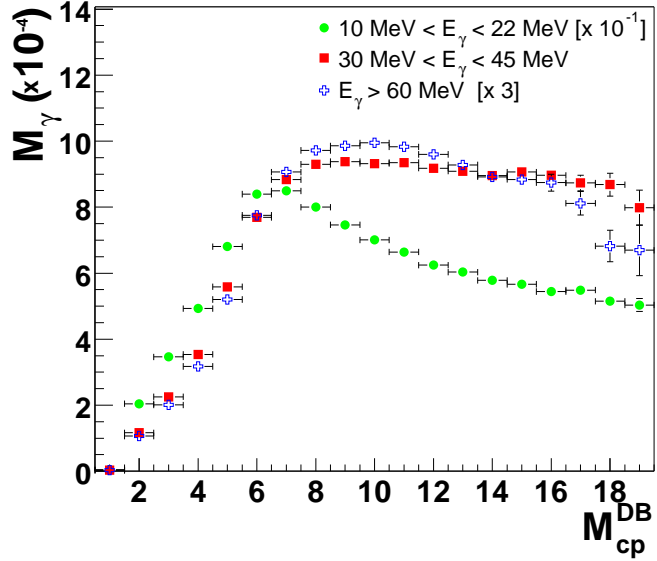}}
 \end{center}
 \caption{\it Experimental photon yield per nuclear reaction,
M$_\gamma$, as a function of the charged-particle multiplicity,
M$_{cp}^{DB}$, measured in the Dwarf-Ball for: (1) GDR photons, (2)
thermal+direct hard-photons, and (3) ``pure" direct hard-photons,
emitted in $^{36}$Ar+$^{197}$Au at 60{\it A} MeV. (The photon
multiplicity distributions in the region of the GDR-photons and of
the direct hard-photons have been scaled to that of the
thermal+direct hard-photons.)}
 \label{fig:enterria:5}
\end{figure}

Thermal and direct bremsstrahlung photons are produced in very
similar reaction channels that are clearly different to those that
lead to GDR statistical photon emission as can be seen in the
distribution of the photon multiplicities (gamma yield per nuclear
reaction) versus the charged-particle multiplicity detected in the
Dwarf-Ball (fig. \ref{fig:enterria:5}). Thermal and direct
hard-photon yields show a very similar dependence with
impact-parameter, increasing a factor $\approx$ 10 when going from
peripheral ($M_{cp}$ = 2) to semi-central ($M_{cp}\,\approx$ 10)
reactions. They saturate in the region of semi-central and central
($M_{cp}$ = 9 - 18) reactions, i.e. at the region of
charged-particle multiplicities corresponding to impact parameters
where the total overlapping of the incident $^{36}$Ar nucleus
inside the much larger $^{197}$Au target nucleus takes place. In
this region of impact-parameters, the mean number of $NN$
collisions saturates, i.e. $\langle N_{pn}
\rangle_b$ = $\langle N_{pn}\rangle_{max}$, and hence so does the
bremsstrahlung photon production. This confirms that
$M_{hard-\gamma}$ is proportional to the number of participants and
ultimately to the volume of the overlap region between projectile
and target at a given impact-parameter.
\\
The GDR photon yield, contrarily, clearly quenches for increasingly
central reactions. Indeed, the yield of photons with 10 MeV $<$
E$_{\gamma}\,<$ 22 MeV is reduced roughly a factor 2 between
$M_{cp}$ = 7 and $M_{cp}\,\approx$ 20. This result confirms the
significant quench of the GDR gamma yield observed experimentally
for increasingly high excitation energies (${\mbox{\large
$\epsilon$}}^{\star}\,>$ 3{\it A} MeV) in several systems
\cite{enterria:Gaar92,enterria:Suom98}. This quenching has been
tentatively interpreted as a result of the loss of collectivity due
to a change from ordered mean-field-driven motion to chaotic
nucleonic motion. This interpretation seems to be nicely
corroborated here.

\section{Summary}

Hard-photon production has been measured in coincidence with most
of the reaction products in four different heavy-ion reactions
($^{36}$Ar+$^{197}$Au, $^{107}$Ag, $^{58}$Ni, $^{12}$C at 60{\it A}
MeV). The first results demonstrate the existence of a thermal
bremsstrahlung component in the three heavier systems showing up in
the region E$_\gamma$ = 30 MeV - 60 MeV as a deviation from the
flatter exponential fall-off describing the high-energy part of the
spectra above 60 MeV. The thermal hard-photon yield accounts for
about 20\% of the total hard-photon yield for the heaviest targets
(Au, Ag and Ni) and is negligible for the $^{36}$Ar+$^{12}$C
system. The thermal slopes scale with the available center-of-mass
energy in the nucleus-nucleus system as expected for a thermal
process taking place after dissipation of the available energy into
internal degrees of freedom over the whole system, whereas direct
slopes scale with the available energy in first-chance $NN$
collisions. Source velocity analysis is consistent with a later
isotropic emission process for the thermal component. The
dependence of the thermal and direct hard-photon yields on impact
parameter is very similar, confirming a common underlying
bremsstrahlung mechanism for their production, significantly
different than that of photons coming from the statistical decay of
collective GDR excitations. The existence of such thermal
bremsstrahlung emission is an unambiguous indication of the
formation of a thermalized nuclear source during the reaction. The
thermal slopes, which are correlated with the temperature of the
excited nuclei \cite{enterria:Schutz97,enterria:Enterria00}, become
a good candidate variable for the determination of the temperature
of the excited nuclear residues remaining after the first
pre-equilibrium phase of the reaction, providing a unique
information of the thermodynamical state of multifragmenting
nuclear systems \cite{enterria:Enterria00}.

\end{document}